\documentclass[physrev,reprint,superscriptaddress,bibnotes]{revtex4-2}  

\usepackage[T1]{fontenc}
\usepackage[utf8]{inputenc} 
\usepackage[australian]{babel}

\usepackage[a4paper,centering,hmargin=1.75cm,vmargin=2cm]{geometry} 
\usepackage{amsmath,amssymb,graphicx,bm,microtype} 
\usepackage[colorlinks,allcolors=blue!50!black]{hyperref}
\usepackage[all]{hypcap}
\usepackage{cleveref,braket,siunitx}
\usepackage[version=4]{mhchem}
\usepackage{cleveref}
\usepackage{tikz}
\usepackage{float}
\usepackage{multirow}
\usepackage{xcolor} 

\let\Re\relax \DeclareMathOperator{\Re}{Re}
\let\Im\relax \DeclareMathOperator{\Im}{Im}

\newcommand{\phis}{\ensuremath{\phi_{\mathrm{s}}}}
\newcommand{\phim}{\ensuremath{\phi_{\mathrm{m}}}}
\newcommand{\phir}{\ensuremath{\phi_{\mathrm{r}}}}
\newcommand{\phib}{\ensuremath{\phi_{\mathrm{b}}}}

\crefname{figure}{Fig.}{Figs.}
\Crefname{figure}{Figure}{Figures}
\crefname{equation}{Eq.}{Eqs.}
\Crefname{equation}{Equation}{Equations}
\crefname{section}{Sec.}{Secs.}
\Crefname{section}{Section}{Sections}
\newcommand{\inlineref}[1]{Ref.~\cite{#1}}
\makeatletter
\AddToHook{cmd/appendix/before}{\def\cref@section@alias{appendix}\def\cref@subsection@alias{appendix}}
\makeatother

\begin{document}

\title{
Programmable quantum simulation of anharmonic dynamics
}

\author{Cameron McGarry}
\email{cameron.mcgarry@sydney.edu.au}
\affiliation{School of Physics, University of Sydney, NSW 2006, Australia}
\affiliation{University of Sydney Nano Institute, University of Sydney, NSW 2006, Australia}

\author{Teerawat Chalermpusitarak}
\affiliation{School of Physics, University of Sydney, NSW 2006, Australia}
\affiliation{University of Sydney Nano Institute, University of Sydney, NSW 2006, Australia}

\author{Kai Schwennicke}
\affiliation{School of Chemistry, University of Sydney, NSW 2006, Australia}

\author{Frank~Scuccimarra}
\affiliation{School of Physics, University of Sydney, NSW 2006, Australia}

\author{Maverick J.~Millican}
\affiliation{School of Physics, University of Sydney, NSW 2006, Australia}

\author{Vassili G.~Matsos}
\affiliation{School of Physics, University of Sydney, NSW 2006, Australia}

\author{Christophe H.~Valahu}
\affiliation{School of Physics, University of Sydney, NSW 2006, Australia}
\affiliation{University of Sydney Nano Institute, University of Sydney, NSW 2006, Australia}
 
\author{Prachi Nagpal}
\affiliation{School of Physics, University of Sydney, NSW 2006, Australia}

\author{Hon-Kwan~Chan}
\affiliation{School of Physics, University of Sydney, NSW 2006, Australia}

\author{Henry L.~Nourse}
\affiliation{School of Chemistry, University of Sydney, NSW 2006, Australia}

\author{Ivan~Kassal}
\affiliation{School of Chemistry, University of Sydney, NSW 2006, Australia}
\affiliation{University of Sydney Nano Institute, University of Sydney, NSW 2006, Australia}

\author{Ting Rei~Tan}
\email{tingrei.tan@sydney.edu.au}
\affiliation{School of Physics, University of Sydney, NSW 2006, Australia}
\affiliation{University of Sydney Nano Institute, University of Sydney, NSW 2006, Australia}

\begin{abstract}
In continuous-variable quantum systems, anharmonicity underlies diverse processes such as proton transfer, false vacuum decay and ferroelectric ordering, which cannot arise in a purely harmonic potential.
Here, we experimentally demonstrate trigonometric-gate-implemented Fourier synthesis (TGIFS), a compilation scheme for quantum simulation of anharmonic dynamics in continuous-variable--discrete-variable (CV--DV) quantum simulators.
TGIFS uses native CV--DV gates, allowing a target anharmonic potential to be determined entirely by classical control parameters, without modification of the hardware.
We programmably implement TGIFS with a single trapped ion, simulating coherent dynamics in a continuously tunable family of double-well potentials and synthesising a quartic phase gate.
The same circuit structure extends to arbitrary potentials and to multiple modes, giving a route to universal programmable simulation of nonlinearly coupled anharmonic potentials as achievable circuit depth increases.
\end{abstract}

\maketitle

Quantum simulation is emerging as a powerful tool for efficiently capturing the dynamics of complex quantum systems that are intractable with classical approaches~\cite{Feynman1982,Lloyd1996, Georgescu2014,Zhang2017,Guo2024, Meth2025, Burov2025,Haghshenas2026, Chalopin2025}.
A promising quantum-simulation paradigm is the direct encoding of continuous-variable (CV) degrees of freedom of the simulated system into controllable oscillators~\cite{Lloyd1996, MacDonell2021, Crane2026, Kemper2025}.
This paradigm can significantly reduce the required quantum resources compared to qubit-only simulations, because discretising CV information into qubits requires deep circuits on many qubits~\cite{Sawaya2020, MacDonell2021, Saha2021, Macridin2022, Richerme2023, Crane2026, Navickas2025, Liu2025Hybrid, Nourse2025}.

Quantum simulators employing CV systems as resources have been used to experimentally simulate chemical dynamics~\cite{Huh2015, Gorman2018, Sparrow2018, McArdle2020, Wang2020, Wang2023, MacDonell2023,Valahu2023,Whitlow2023,So2024, Navickas2025, Sun2025,Nielsen2025}, high-energy physics~\cite{Gerritsma2010, Gerritsma2011} and quantum field theory~\cite{Than2025, Saner2026}.
There are also proposals to use them for further study of quantum field theories~\cite{Davoudi2021, Crane2026, Liu2025Hybrid, Araz2025}, condensed matter physics~\cite{Liu2025Hybrid, Kemper2025} and quantum chaos~\cite{Gardniner1997}.

A central challenge in CV-based simulation is capturing anharmonic dynamics.
Quantum simulators typically use high-quality harmonic CV degrees of freedom.
Hence, simulating anharmonicity requires realising non-Gaussian operators, which rely on nonlinear interactions that are typically weak and therefore challenging to implement.
Nevertheless, the anharmonic regime is important in chemistry, because many molecular vibrations are anharmonic~\cite{Bender1969, Silbey1980, Miller1983, Munn1985, Brack1993, Hamm2012, Zhu2024, Mallweger2026}, as well as in quantum field theory~\cite{Faddeev1978} and quantum many-body systems~\cite{Stewart1968, McCumber1968, Coleman1975}.

Therefore, a crucial feature of multi-purpose CV simulators is the programmable implementation of anharmonicity.
Programmability is the ability to choose a target Hamiltonian by varying experimental parameters without modification of the underlying hardware.
This would enable high-throughput quantum simulations without the need to build Hamiltonian-specific hardware.

Existing approaches to simulating anharmonic dynamics have not demonstrated fully programmable anharmonicity.
For instance, simulations of anharmonicity with photonic systems are limited by weakly nonlinear optical elements~\cite{Sparrow2018, Nielsen2025}.
Tailored superconducting circuit quantum electrodynamics devices have been used to simulate dissipative dynamics in tunable double-well potentials~\cite{deAlbornoz2024}.
There, the hardware-intrinsic Kerr nonlinearity restricts programmability over arbitrary forms of anharmonicity.
Similarly, trapped-ion devices allow implementing specific anharmonic dynamics through customised devices~\cite{Brown2011,Harlander2011,Tanaka2021,Niedermeyer2025} or engineered light-atom couplings~\cite{Leibfried2002,Bazavan2026,Simoni2025arXiv}.
However, all three platforms lack an efficient compilation layer between target evolution operators and experimentally accessible gates, which limits the anharmonic potentials and non-Gaussian gates that can be implemented.

Bosonic quantum signal processing (BQSP) presents a route to programmable quantum simulation of anharmonic dynamics.
BQSP operates on hybrid systems composed of both CV and discrete variable (DV) degrees of freedom, such as qubits.
It uses CV--DV primitives---qubit-state-dependent displacements (SDD) and single-qubit rotations (SQR)---which are interleaved to realise a nonlinear transformation on the CV subsystem.
While universal control of CV--DV simulators has long been established~\cite{Law1996, Leibfried2002, Walschaers2021, Sutherland2021, Matsos2024}, BQSP allows the construction of non-Gaussian operators in a systematic fashion with a well-defined circuit structure and error scaling by appropriately choosing amplitudes and phases for the primitive operations~\cite{Eickbusch2022, Park2024, Sinanan-Singh2024, Liu2025, Hong2026, Fong2026, TCKS, Rainaldi2026, Laneve2025}.
Furthermore, BQSP techniques can construct an operator that encodes the time evolution under a target Hamiltonian, enabling simulation of dynamics.

\begin{figure}
    \centering
   \includegraphics[width=\columnwidth]{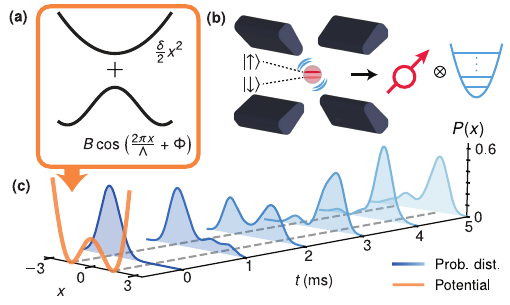}
    \caption{
    \textbf{Quantum simulation of dynamics in an anharmonic potential.}
    \textbf{(a)}~A double-well potential is synthesised using a parabolic contribution and a cosine contribution from a trigonometric gate.
    \textbf{(b)}~A trapped-ion quantum simulator, with the spin providing the qubit and the motion the oscillator.
    \textbf{(c)}~Dynamics in a symmetric double well; a wavepacket initialised in one well tunnels through the barrier to the opposite well.
    }
    \label{fig:1}
\end{figure}

Here, we experimentally demonstrate programmable quantum simulation of anharmonic dynamics.
We do so through an efficient compilation layer, trigonometric-gate-implemented Fourier synthesis (TGIFS), which uses BQSP to compile target evolution operators into programmable non-Gaussian gate sequences.
TGIFS provides programmability for general anharmonic dynamics and non-Gaussian gates, which are defined entirely by classical experimental control parameters.
We perform our experiments using a single trapped ion, and implement the gates through light-atom interactions. 
We demonstrate the capabilities of TGIFS by simulating a family of double-well potentials (\cref{fig:1}) and implementing the quartic phase gate.

\section{Simulating anharmonicity}
\label{sec:sim_anhm}

We simulate quantum dynamics in anharmonic potentials using TGIFS~\cite{TCKS}, which describes simulation of the Hamiltonian
\begin{equation}
H = \frac\delta2 (x^2+p^2)+V(x),
\label{eq:H_general}
\end{equation}
where $\delta$ is the frequency of the oscillator's harmonic component, $x = (a + a^\dagger)/\sqrt{2}$ and $p = -i(a-a^\dagger)/\sqrt{2}$ are the dimensionless position and momentum operators, respectively, and $V(x)$ is a time-independent anharmonic potential.
In TGIFS, $V(x)$ is decomposed into a Fourier series whose terms are individually synthesised using a set of non-Gaussian operators.
We consider the one-dimensional case (i.e., a single mode), but TGIFS can be straightforwardly extended to multiple modes, allowing one to simulate any potential that is a function of mutually commuting oscillator quadratures~\cite{TCKS}.

We simulate the dynamics generated by $H$ in the interaction picture with respect to the harmonic term.
In this picture, the dynamics is governed by
\begin{align}
    H_\mathrm{I}(t) &= U_0^\dagger(t)HU_0(t),\\
    &= V(x_\mathrm{I}(t)),
\end{align}
where $U_0(t) = \exp(-i \tfrac\delta2(x^2+p^2) t)$ and $x_\mathrm{I}(t) = (ae^{-i\delta t} +a^\dagger e^{i\delta t})/\sqrt{2}$ is the time-dependent position operator.
The goal is to simulate the unitary evolution generated by $H_\mathrm{I}(t)$ from $t_0$ to $T$, 
\begin{equation}
U(t_0, T) = \mathcal{T}_+ \exp\bigg(-i\int_{t_0}^T
H_\mathrm{I}(t)\,\mathrm{d} t\bigg),
\end{equation}
where $\mathcal{T}_+$ is the time-ordering operator.
We approximate this integral as a product over small timesteps $\Delta t$ (so that $\delta \Delta t \ll 1$) and take $H_\mathrm{I}$ to be constant throughout each step, so that
\begin{equation}
U(t_0, T)\approx\prod_{k=0}^{K-1} e^{-iV(x_{\mathrm{I}}(k\Delta t)) \Delta t},
\label{eq:time_trott}
\end{equation}
where $K=T/\Delta t$; we assume that, during the $k$th timestep, $H_\mathrm{I}(t)$ is time independent and equals $V(x_\mathrm{I}(k\Delta t))$.

To enact time evolution at step $k$, $V(x_\mathrm{I}(k\Delta t))$ is expanded into a truncated Fourier series, whose terms correspond to the gates that synthesise the potential.
We assume the dynamics is confined to the region of interest (ROI) $x\in[-\ell/2,\ell/2]$.
We expand $V$ over this interval in a Fourier series,
\begin{align}
    V(x_\mathrm{I}(k\Delta t))\approx\sum_{n=1}^{N} B_n \cos\bigg(\frac{2\pi n}{\Lambda}x_\mathrm{I} (k\Delta t)+\Phi_n\bigg),
    \label{eq:FTpot}
\end{align}
where $N$ is the number of Fourier terms, $\Lambda=\ell$ is the period, and $B_n$ and $\Phi_n$ are the Fourier coefficients. 
The dynamics at the $k$th timestep is described by the operator
\begin{align}
     U(k\Delta t, (k+1)\Delta t) &= e^{-iV(x_\mathrm{I}(k\Delta t))\Delta t}\\
    &\approx \prod_{n=1}^{N}G_\mathrm{c}^{(n)}(k\Delta t),
    \label{eq:step_evo}
\end{align}
where we have approximated the unitary as a sequence of operators, and the approximation becomes exact in the limit $N\rightarrow\infty$ and $\Delta t\rightarrow0$.
$G_\mathrm{c}^{(n)}$ is the \emph{cosine trigonometric gate}~\cite{TCKS}
\begin{equation}
    G_\mathrm{c}^{(n)}(k\Delta t) = \exp\left(-i\Delta t B_n\cos\left(\frac{2\pi nx_\mathrm{I}(k\Delta t)}{\Lambda} + \Phi_n\right)\right).
    \label{eq:trig_gate}
\end{equation}
Thus, the task of simulating any potential $V(x)$ of \cref{eq:H_general} amounts to implementing a series of cosine trigonometric gates.
To simulate a single timestep, $G_\mathrm{c}^{(n)}$ is applied $N$ times, with the parameters of the gate corresponding to the parameters of the Fourier expansion in \cref{eq:FTpot} ($B_n$, $\Phi_n$ and $\Lambda$).

\begin{figure*}
    \centering
   \includegraphics[width=\textwidth]{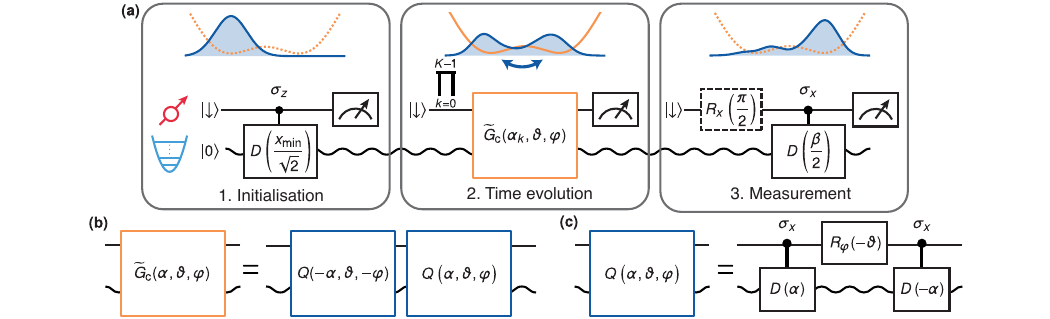}
    \caption{
    \textbf{Circuit diagram for simulating dynamics in double-well potentials.}
    \textbf{(a)} The experiment consists of three steps.
    1.~Initialisation of the wavepacket at a chosen position. 
    The potential is a symmetric double well which is not synthesised at this stage (dotted line).
    The initial position is the minimum of the potential at $-x_\mathrm{min}$.
    2.~Time evolution of the wavepacket in the double-well potential (solid line) is simulated using the trigonometric gate, $\widetilde{G}_\mathrm{c}$. 
    3.~Measurement of the oscillator state. 
    Dashed box: optional operation to extract the imaginary part of the characteristic function.
    \textbf{(b)} The trigonometric gate is constructed with two BQSP subroutines.
    \textbf{(c)} The BQSP subroutine consists of two SDDs and an SQR. 
    }
    \label{fig:circuit}
\end{figure*}

Our implementation of $G_\mathrm{c}^{(n)}(k\Delta t)$ is shown in \cref{fig:circuit}b--c; it consists of a BQSP sequence, which is Trotterised to isolate the targeted cosine operator.
We will now show that this sequence yields a cosine trigonometric gate.
We begin with the CV--DV primitives that form the BQSP sequence, which consist of a state-dependent displacement (SDD) 
\begin{equation}
    D(\sigma_x\alpha) = \exp\left(\sigma_x(\alpha a^\dagger  - \alpha^*a)\right)
\end{equation}
and a single-qubit rotation (SQR)
\begin{equation}
    R_\varphi(\vartheta)= \exp(-\tfrac12i\vartheta\,\hat{\mathbf n}(\varphi)\cdot\boldsymbol{\sigma}),
\qquad
\label{eq:R_phi}
\end{equation}
where the complex number $\alpha=|\alpha|e^{i(\zeta - \pi/2)}$ is given in terms of its magnitude, $|\alpha|$, and quadrature angle $\zeta \in[0,2\pi)$;  $\vartheta$ is the angle of rotation, $\mathbf{\hat{n}} = (0, \sin\varphi, \cos\varphi)$ is the axis of rotation; and $\boldsymbol{\sigma} = (\sigma_x, \sigma_y, \sigma_z)$ is the vector of Pauli operators. 

We construct a BQSP sequence~\cite{Laneve2025} by placing an SQR operator between two SDDs,
\begin{equation}
    Q(\alpha, \vartheta, \varphi) =  D(-\sigma_x\alpha) R_\varphi(-\vartheta) D(\sigma_x\alpha),
    \label{eq:Q_def}
\end{equation}
which can be rewritten (see \cref{appendix:trig_gates}) as
\begin{multline}
    Q(\alpha, \vartheta, \varphi) = \exp\big(
    \tfrac12 i\vartheta\big(\sigma_z\cos(2\sqrt{2}|\alpha| x_\zeta + \varphi) \\
    \hspace{2cm}+\sigma_y\sin(2\sqrt{2}|\alpha| x_\zeta +\varphi)\big)\big),
    \label{eq:Q}
\end{multline}
where $x_\zeta = (ae^{-i\zeta} + a^\dagger e^{i\zeta})/\sqrt{2}$ is the rotated quadrature at phase $\zeta$.

\Cref{eq:Q} contains the desired cosine operator for implementing $G_\mathrm{c}^{(n)}$, as well as a term dependent on sine.
We isolate the cosine term using the first-order Trotter decomposition~\cite{Trotter1959}
\begin{equation}
e^{i(F+G)r} = e^{iFr}e^{iGr} + \mathcal{O}(r^2),
\end{equation}
where $F$ and $G$ are operators and $r$ is a small parameter.
We set $e^{iFr}= Q(\alpha, \vartheta, \varphi)$, $e^{iGr}=Q(-\alpha, \vartheta, -\varphi)$ and take $\vartheta = r$, yielding a new operator which is the BQSP-implemented trigonometric gate conditioned on $\sigma_z$,
\begin{align}
   \widetilde{G}_\mathrm{c}(\alpha, \vartheta, \varphi) &= Q(\alpha,\vartheta, \varphi)Q(-\alpha, \vartheta, -\varphi) \label{eq:cosine_def} \\
   &= \exp(i\vartheta \sigma_z\cos(2\sqrt{2}|\alpha| x_\zeta + \varphi)) + \mathcal{O}(\vartheta^2).
   \label{eq:cosine_evo}
\end{align}

With an appropriate choice of parameters ($\alpha, \vartheta, \varphi$), $\widetilde{G}_\mathrm{c}$ realises the programmable target gate $G_\mathrm{c}^{(n)}$.
The qubit-dependence of $\widetilde{G}_\mathrm{c}$ is removed by first preparing the qubit in an eigenstate of $\sigma_z$.
Throughout this work, we initialise the qubit in $\ket{\downarrow}$.
In particular, we implement
\begin{align}
G_\mathrm{c}^{(n)}(k\Delta t) = \bra{\downarrow}\widetilde{G}_\mathrm{c}(\alpha,\vartheta,\varphi)\ket{\downarrow},
\label{eq:G_ideal_approx}
\end{align}
by setting 
\begin{align}
    \alpha = \frac{\pi n}{\sqrt{2}\Lambda} e^{i(k \delta \Delta t - \pi/2)}, &&
    \vartheta = B_n \Delta t, &&
    \varphi = \Phi_n,
    \label{eq:Gc_params_fourier}
\end{align}
and neglecting the $\mathcal{O}(\vartheta^2)$ term.

In the following sections, we use TGIFS to simulate the evolution of wavepackets in double-well potentials (see \cref{fig:1}).
Our target Hamiltonian is
\begin{equation}
H_\text{sim} = \frac{\delta}{2}(x^2 + p^2) + B\cos\bigg(\frac{2\pi x}{\Lambda} + \Phi\bigg),
\label{eq:H_sim}
\end{equation}
which can be synthesised using TGIFS with a single Fourier component ($N=1$).
The corresponding potential is the sum of a harmonic term and a single trigonometric component, which yields a double-well landscape near the origin.
In this case, the evolution operator for the $k$th timestep is $\widetilde{G}_\mathrm{c}(\alpha_k, \vartheta, \varphi)$, where
\begin{align}
    \alpha_k=\frac{\pi}{\sqrt{2}\Lambda}e^{i(\zeta-\pi/2)}, &&
    \zeta = k \delta \Delta t, &&
    \vartheta = B \Delta t, &&
    \varphi = \Phi.
    \label{eq:Gc_params}
\end{align}
Each of $|\alpha|$, $\zeta$, $\vartheta$, and $\varphi$ can be experimentally adjusted, as detailed in \cref{sec:experiment}, making the simulation programmable.

\section{Methods}
\label{sec:methods}

\subsection{Experimental implementation}
\label{sec:experiment}

We simulate the dynamics under $H_\text{sim}$ using a trapped-ion system comprising an \ce{^{171}Yb^+} ion confined in a Paul trap~\cite{Navickas2025, Matsos2025}. 
The bosonic (CV) information is encoded in one of the radial modes of motion, which has secular frequency $\omega_x = 2\pi\times \SI{1.39}{\mega\hertz}$. 
The motional mode has negligible native anharmonicity due to the large ion-electrode distance (\SI{565}{\micro\meter})~\cite{Home2011}. 
Dephasing of the motional state occurs at a rate of $\gamma_\phi = \SI{18}{\per\second}$; the heating rate of \SI{0.2}{quanta~s^{-1}}~\cite{MacDonell2023} is comparatively insignificant~\cite{Valahu2023}. 
The DV information is encoded in the hyperfine clock states of the $^2\mathrm{S}_{1/2}$ level, with a qubit frequency $\omega_0 = 2\pi\times\SI{12.6}{\giga\hertz}$. The qubit has a measured $T_2^*$ time of \SI{8.7}{\second}~\cite{Tan2023}, providing excellent coherence as a nonlinear element to engineer high-quality anharmonicity in the CV system. 

Both the qubit and the oscillator are coherently controlled via stimulated Raman transitions, driven by two laser beams derived from a $\lambda=\SI{355}{\nano\meter}$ pulsed laser.
The frequency, amplitude, phase, and duration of the two beams are modulated independently by acousto-optic modulators.
We operate in the Lamb--Dicke regime, $\eta^2(2\bar{n}+1)\ll 1$, where $\bar{n}$ is the average occupation number of the oscillator, which does not exceed 1.2 throughout the coherent evolution, and $\eta=2\pi x_\mathrm{gnd}/\lambda=0.08$ is the Lamb--Dicke parameter.
Here $x_\mathrm{gnd} = \sqrt{\hbar/(2m\omega_x)} = \SI{4.6}{\nano\meter}$ is the spatial extent of the oscillator's ground-state wavefunction, where $m$ is the ion's mass.
Operation in the Lamb--Dicke regime allows driving sideband-resolving spin-motion interactions and neglecting higher-order ($\eta^2$ and higher) qubit-oscillator couplings.

Control of the qubit is achieved by tuning the frequency difference between the two beams to the qubit frequency.
The interaction Hamiltonian after making the rotating-wave approximation is
\begin{equation}
    H_\text{Q} =
    \tfrac12 \Omega_0\sigma_x,
\end{equation}
where $\Omega_0$ is the Rabi frequency.
To realise the SQR $R_\varphi(\vartheta)$, we decompose it into three primitive rotations:
\begin{equation}
    R_\varphi(\vartheta)=R_x(-\varphi)R_z(\vartheta)R_x(\varphi),
    \label{eq:R_phi_xzx}
\end{equation}
where $R_m(\theta)=\exp(-i\theta\sigma_m/2)$ for $m\in\{x,y,z\}$. 
$R_x(\varphi)$ is obtained by applying $H_\text{Q}$ for time $\tau_\varphi=\varphi/\Omega_0$,
\begin{equation}
R_x(\varphi) = e^{-iH_\mathrm{Q}\tau_\varphi}.
\end{equation}
Arbitrary rotations in $x$ and $y$ can be applied with other choices of times and phases. 
$R_z(\vartheta)$ rotations are implemented by shifting the phase of the qubit’s reference local oscillator during the SQR pulse and for all subsequent operations.

Coherent qubit-oscillator operations are driven by tuning the frequency difference between the two Raman beams close to the motional sidebands of the carrier transition.
We drive red-sideband (RSB) and blue-sideband (BSB) interactions,
\begin{align}
H_{\mathrm{RSB}} &= \tfrac12 i\Omega \big(
a\sigma_+e^{i(\delta_\mathrm{L} t - \phir)} \big)+ \text{h.c.}\\
H_{\mathrm{BSB}} &= -\tfrac12 i\Omega \big(a\sigma_-e^{i(\delta_\mathrm{L} t + \phib)} \big)+ \text{h.c.},
\end{align}
where $\Omega=\eta\Omega_0$ is the Rabi frequency of the sideband transitions, $\delta_\mathrm{L}$ is the laser detuning from the motional sideband frequency, $\sigma_\pm = (\sigma_x \pm i\sigma_y)/2$ are the qubit raising ($+$) and lowering ($-$) operators, and $\phir$ and $\phib$ are the relative phases of the light for the RSB and BSB, respectively.
Driving both interactions simultaneously couples the spin to the motion with the Hamiltonian
\begin{align}
    H_\text{SDD} &= H_\text{RSB} + H_\text{BSB} \\
    &=-\frac{\Omega}{2} \sigma_{\phis} \big( ae^{i(\phim + \delta_\mathrm{L} t )} \big)+ \text{h.c.} ,
    \label{eq:H_sdd_main}
\end{align}
where $\sigma_{\phis} = \sigma_x\cos\phis + \sigma_y\sin\phis$, $\phis=(\phir+\phib)/2$ and $\phim=(\phib-\phir)/2$ are the qubit and the oscillator phases, respectively.

By setting $\delta_\mathrm{L}=0$, $\phis=0$ and $\phim=\pi/2$, driving $H_\text{SDD}$ for a duration $\tau$ implements the SDD
\begin{equation}
e^{-iH_\text{SDD}\tau} = D\big(\tfrac12\sigma_x\Omega\tau \big).
\label{eq:SDF_SDD}
\end{equation}

\subsection{Pulse sequence}

The experiment simulates the dynamics of a wavefunction within a double-well potential by implementing the above control and by performing tomography to reconstruct the motional wavepacket throughout its evolution.

We set the parameters of the target Hamiltonian $H_\mathrm{sim}$ to resolve tunnelling dynamics within experimentally accessible operating conditions.
Initially, we investigate a symmetric double well, with minima located at $\pm x_\mathrm{min}=\pm1.5$.
We set $|\alpha|=\pi/6$, $\delta=2\pi\times\SI{500}{\hertz}$, $\vartheta=0.8$, and $\varphi=0$, which yields a symmetric double-well potential, and the barrier height is $V(0)-V(x_\mathrm{min})=0.93\delta$.

The experiment is separated into three steps, depicted in \cref{fig:circuit}: \textit{initialisation} of the wavepacket at the desired location; \textit{time evolution} using the trigonometric gate to simulate dynamics; and \textit{measurement} of the final state.

\subsubsection{Initialisation}
\label{sec:methods:init}

We first prepare the oscillator in the ground state with a mean occupation of $0.04$ quanta by Doppler cooling followed by sideband cooling.
The qubit is prepared in $\ket{\downarrow}$ by optical pumping.
See \cref{appendix:state_prep} for details.

Next, we displace the wavepacket from the origin to the desired initial position at the centre of the left well (at $-x_\mathrm{min}$).
To do this, we implement a three-pulse sequence.
We first apply a rotation $R_y(-\pi/2)$ to transfer the qubit into the $\ket{+}$ eigenstate of $\sigma_x$.
Next, we apply the SDD $D(-\sigma_xx_\mathrm{min}/\sqrt{2})$.
Finally, $R_y(\pi/2)$ returns the qubit to $\ket{\downarrow}$.

We apply a mid-circuit measurement on the qubit to reduce erroneous spin flips during state preparation (see \cref{appendix:measurement}).
A measurement outcome of $\ket{\uparrow}$ implies an error in the initialisation procedure, and the result of the experiment is ignored.
Our initialisation procedure measures a successful $\ket{\downarrow}$ outcome on $>\SI{99}{\percent}$ of experimental shots, a value comparable with our state-preparation and measurement fidelities.

\subsubsection{Time evolution}
\label{methods:time_evo}

Following initialisation, we apply the trigonometric gate $\widetilde{G}_\mathrm{c}$ $K$ times to implement stepwise evolution under $H_\mathrm{sim}$ for total duration $K\Delta t$.
The state $\ket{\psi(0)}$ evolves according to
\begin{equation}
    \ket{\psi(K\Delta t)} = \left(\prod_{k=0}^{K-1} \widetilde{G}_\mathrm{c}(\alpha_k,\vartheta, \varphi)\right)\ket{\psi(0)},
    \label{eq:state_evo}
\end{equation}
with parameter values given in \cref{eq:Gc_params}.
We implement $\widetilde{G}_\mathrm{c}$ using the sequence of SQR and SDD operations shown in \cref{fig:circuit}. We set $\phis=0$ and $\phim=\pi$, so that $\zeta=\pi-\phim-\delta_\mathrm{L}t$.
To realise the given $|\alpha|$ and $\varphi$, we choose the SDD pulse duration $\tau$ and the SQR duration $\tau_\varphi$ so that
\begin{align}
    \tau = \frac{2|\alpha|}\Omega &&  \text{and} && \tau_\varphi =\frac\varphi\Omega_0,
    \label{eq:pulse_durations}
\end{align}
where Rabi frequencies $\Omega_0=\pi/\SI{35}{\micro\second}$ and $\Omega=\pi/\SI{150}{\micro\second}$ are determined by the laser intensity at the ion.
Hence, for the targeted double-well potential with parameters stated above, we choose $\tau=\Delta t/4=\SI{50}{\micro\second}$ and $\tau_\varphi=0$.
$\vartheta$ is realised using an $R_z$ rotation, as described in \cref{sec:experiment}.
The phase $\zeta=\delta k\Delta t$ in $\widetilde{G}_\mathrm{c}$ increases with the timestep $k$.
This increase is implemented through $H_\mathrm{SDD}$ (\cref{eq:H_sdd_main}) where we set $\phim$ to be constant and $\delta_\mathrm{L}=-\delta=-2\pi\times\SI{500}{\hertz}$.

After time evolution, we perform a second mid-circuit measurement, and post-select on $\ket{\downarrow}$ outcomes.
The derivation of $H_\mathrm{sim}$ in \cref{sec:sim_anhm} assumed that the qubit remains in $\ket{\downarrow}$ throughout the evolution. 
However, Trotter errors in \cref{eq:cosine_evo} can lead to spin flips.
By rejecting spin-flip outcomes, the mid-circuit measurement mitigates these errors.

\subsubsection{Measurement}
\label{methods:meas}

The state of the oscillator is measured at various points throughout the evolution under the double-well potential.
The motional state is measured by first entangling with the qubit followed by projective qubit measurements. 
We extract the characteristic function $\chi(\beta)=\langle D(\beta)\rangle$, where the expectation value is taken with respect to the motional state~\cite{Fluhmann2020}.
This approach can be used for both full tomography (allowing the reconstruction of the state's Wigner function, $W(x,p)$) as well as measurements that directly give information such as the probability distributions $P(x)$ or the expectation values of quadratures.

The measurement protocol is depicted in \cref{fig:circuit}a.
We measure $\chi(\beta)$ by applying the SDD $D(\sigma_x\beta/2)$, followed by a qubit measurement in the computational ($\sigma_z$) basis (see \cref{appendix:measurement}).
The expectation value of the measurement gives the real part of the characteristic function, $\langle\sigma_z\rangle=-\Re[\chi(\beta)]$.

Prepending $R_x(\pi/2)$ (dashed box) gives the imaginary part, $\langle\sigma_z\rangle=-\Im[\chi(\beta)]$.
These expectation values are estimated by repeating the measurement procedure and averaging the results.
This is repeated for all desired values of $\beta$.
Due to the Hermitian symmetry $\chi(-\beta) = \chi(\beta^*)$, scanning half of the phase space is sufficient to fully construct $\chi(\beta)$.

\begin{figure}
    \centering
    \includegraphics[width=\columnwidth]{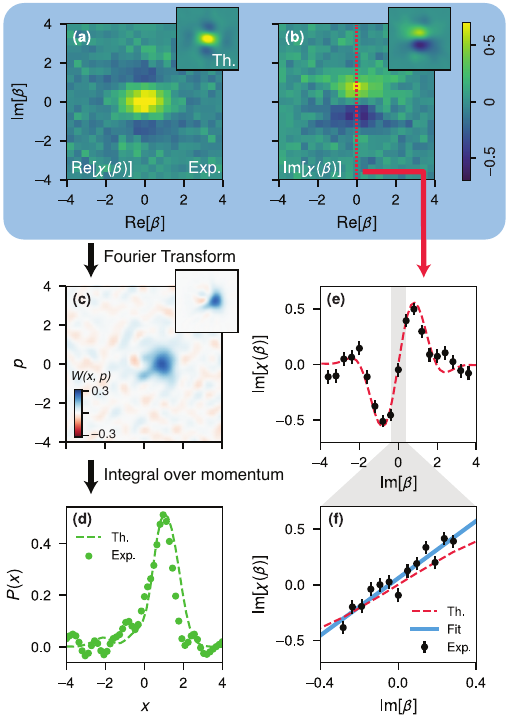}
    \caption{
    \textbf{Reconstruction of motional wavepackets and extraction of position expectation values}. Given are representative measurements at $t=\SI{4}{\milli\second}$ for a symmetric double well.   
    \textbf{(a,b)}~Real and imaginary parts of the characteristic function $\chi(\beta)$.
    \textbf{(c)}~Wigner function $W(x,p)$ obtained as the Fourier transform of $\chi(\beta)$.
    \textbf{(d)}~Probability distribution $P(x)$, obtained as $\int dp\,W(x,p)$.
    \textbf{(e)}~Slope of $\Im[\chi(\beta)]$ at $\Im[\beta]=0$ determines $\langle x\rangle$ (\cref{eq:xexpect}).
    \textbf{(f)}~Finer measurement of $\partial\Im[\chi(\beta)]/\partial\Im[\beta]$ to determine $\langle x\rangle$.
    All theoretical models include dephasing and Trotter error.
    }
    \label{fig:tomography}
\end{figure}

\begin{figure*}
    \centering
   \includegraphics[width=\textwidth]{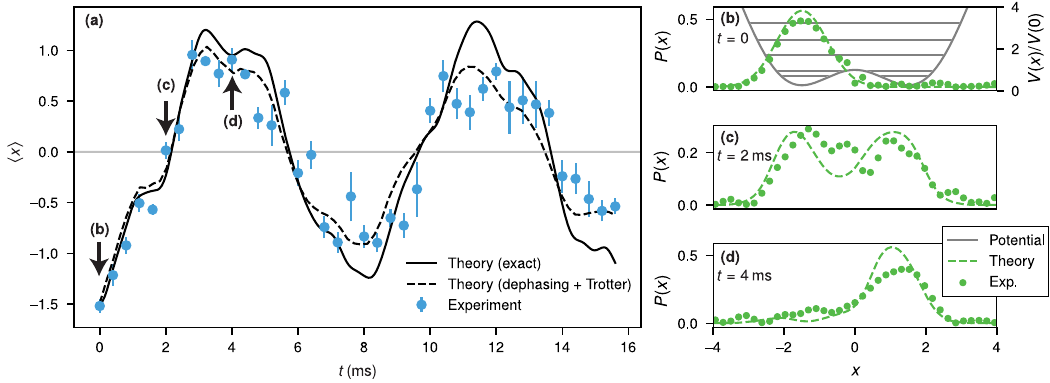}
    \caption{\textbf{Quantum tunnelling in a synthesised symmetric double-well potential.}
    \textbf{(a)} Expectation value $\langle x \rangle$ of the wavepacket's position through time, showing oscillations due to tunnelling through the barrier centred at $x=0$. 
    Experimental points are determined as in \cref{fig:tomography}f, with error bars obtained from the uncertainty in the slope of the fit line.
    We also show the theoretical predictions for the exact dynamics under $H_\mathrm{sim}$ (solid line) and including dephasing and Trotter error (dashed line).
    \textbf{(b--d)} Reconstructed probability distributions $P(x)$ at $t=\SI{0}{\milli\second}$, $t=\SI{2}{\milli\second}$ and $\SI{4}{\milli\second}$, compared with theoretical predictions (including post-selection, dephasing and Trotter error).
    Also shown in (b) is the target potential and the associated energy levels.
    The combined probability of occupying the two lowest levels (both below the barrier) is \SI{95}{\percent}.
    }
    \label{fig:symmetric}
\end{figure*}

Representative results of this measurement are shown for a symmetric double well in \cref{fig:tomography}a--b at $t=\SI{4}{\milli\second}$, which corresponds to the time it takes the wavepacket to tunnel to the opposite side of the barrier.
A Fourier transform of $\chi(\beta)$ gives the Wigner function (\cref{fig:tomography}c)
\begin{equation}
W(x, p) = \frac{1}{2\pi^{2}} \int \mathrm{d}^{2}\beta \, \chi(\beta) e^{\gamma \beta^{*} - \gamma^{*} \beta},
\label{eq:WignerFT}
\end{equation}
where $\gamma=(x+ip)/\sqrt{2}$.
The probability distribution of the state can be obtained through the integral of $W(x,p)$ along the orthogonal axis. For example, the position probability distribution is given by,
\begin{equation}
P(x) = \int_{-\infty}^\infty W(x,p)\,\mathrm{d}p,
\end{equation}
which is shown in \cref{fig:tomography}d.

The distribution $P(x)$ can also be reconstructed from fewer measurement samples by performing a one-dimensional scan of $\chi(\beta)$ with $\Re[\beta]=0$, 
\begin{equation}
\label{eq:Pfromchi}
P(x) =\frac{1}{2\pi}\int_{-\infty}^{\infty} du\; e^{-iux}\; \chi\!\left(\frac{i u}{\sqrt{2}}\right).
\end{equation}
In this approach, the number of measurements of $\chi(\beta)$ is reduced compared to integrating $W(x,p)$.

The expectation value of a quadrature can be extracted with even fewer measurements by estimating the slope of the characteristic function at the origin~\cite{Gerritsma2010}.
In particular, $\langle x \rangle$ is given by~\cite{Gerry2004},
\begin{equation}
\langle x\rangle = \frac{1}{\sqrt{2}}\frac{\partial \Im[\chi(\beta)]}{\partial\Im[\beta] }\bigg|_{\beta=0}.
\label{eq:xexpect}
\end{equation}
In our measurement, we fit a line to experimental data points to determine $\langle x\rangle$ (see \cref{fig:tomography}f).

We can also extract $\langle x\rangle$ by measuring only one value of $\chi(\beta)$ and 
finding the slope at the origin from two-point finite difference (2PFD).
For the two points, we take the first to be at the origin, $\Im[\chi(0)]=0$ (because $\chi(0)=1$), and the second to be $\Im[\chi(ih)]$.
When $h$ is sufficiently small, $\chi$ can be treated as linear, and the expectation value is found by approximating~\cref{eq:xexpect} as
\begin{equation}
    \langle x\rangle\approx\frac{\Im[\chi(ih)]}{\sqrt{2}h}. 
    \label{eq:xapprox}
\end{equation} 
This scheme requires a suitable choice of $h$: if $h$ is too small, the quantum projection noise will be greater than $\Im[\chi(ih)]$ (for a given number of measurements) and if $h$ is too large, error arises due to nonlinearity in $\chi(\beta)$.
We use $h=0.4$ throughout this work, which was chosen because it ensures the total error from shot noise and the 2PFD is small compared to other errors, as shown in \cref{appendix:error}.

\section{Results}
\label{results}

The coherent and fully tunable control afforded by TGIFS opens new opportunities to study anharmonic potentials across the physical sciences, as we demonstrate here in three stages.
First, we simulate tunnelling dynamics in a symmetric double-well potential; we observe a wavepacket coherently traversing the barrier four times.
Second, we tune the experimental parameters to programmably introduce asymmetry, showing the ability to explore a family of double-well potentials without changing the underlying hardware.
Finally, we implement the quartic phase gate using $N=2$ Fourier terms, demonstrating TGIFS's capability to simulate arbitrary potentials using the multi-term Fourier synthesis offered in \cref{eq:step_evo}.

We choose the double-well potential as an example because it is prevalent across the physical sciences.
In physics, asymmetric double-well potentials describe false vacuum decay~\cite{Coleman1977, Zenesini2024} and set the ordering temperature of  hydrogen-bonded ferroelectrics~\cite{Koval2002}.
In chemistry, tunnelling opens reaction pathways~\cite{Wild2023}, sets isomerisation rates~\cite{Cribb1982} and, in asymmetric potentials, may affect proton transfer in hydrogen-bonded systems~\cite{Godbeer2015, Slocombe2022}.
Tunnelling is visible in the spectra of inverting molecules~\cite{Bouakline2021, Nguyen2021, Nguyen2025}, and of strongly anharmonic crystals~\cite{Mitoli2025}.
In many of these settings, the one-dimensional double well is a simplification of a multimode problem, which is made so as to render the dynamics classically tractable~\cite{Bouakline2021} or comparable to exact results~\cite{Shuryak1988}.

The quartic phase gate---$e^{-i\kappa x^4}$, where $\kappa$ is a parameter---provides a fundamental building block for modelling nonlinear dynamics.
The quartic term represents the lowest-order even anharmonicity of an oscillator and appears as a first-order correction in molecular vibrational perturbation theory~\cite{Nielsen1951, Barone2005}.
In quantum optics, its number-conserving part contains a Kerr interaction, ${a^\dagger}^2 a^2$, which can be approximated by interleaving quartic phase gates at rotated quadrature angles (see \cref{appendix:Kerr}).
Kerr interactions produce nonlinear dynamics such as the collapse and revival of coherence~\cite{Greiner2002}.
They also provide on-site interaction of Bose--Hubbard systems~\cite{Fisher1989, Jaksch1998}, where their competition with tunnelling can produce nonlinear bosonic Josephson junctions and interaction-induced self-trapping~\cite{Raghavan1999, Abbarchi2013}.
Recently, BQSP circuits were also used to synthesise the cubic phase gate---$e^{-i\kappa x^3}$---in a superconducting device~\cite{Fontaine2026} and a trapped ion~\cite{LeeJunsheng2026}.

\subsection{Symmetric double well}
\label{symmetric}

Wavepacket tunnelling in the symmetric double well behaves as expected, see \cref{fig:symmetric}a.
The wavepacket is initialised in the left well as described in \cref{sec:methods:init}.
As time evolves, the slow oscillation of $\langle x\rangle$ indicates that the wavepacket tunnels through the barrier four times in \SI{16}{ms}. 
We observe good agreement between the experiment and the theory.
In particular, we observe only partial revivals, that is, the wavepacket does not return to its initial spatial profile because of the population in states with energies above the barrier.

\Cref{fig:symmetric}b--d shows the reconstructed wavepacket probability distributions $P(x)$ at three times, revealing the localisation of the wavepacket in each well at the turning points of the oscillation ($t=0$ and $t=\SI{4}{\milli\second}$), as well as its delocalisation across the two wells at an intermediate time ($t=\SI{2}{\milli\second}$).
\Cref{appendix:motional_meas} gives the complete data for $\chi(\beta)$ used to construct these probabilities, as well as further validation by scanning the entire characteristic function.
Results for $t=\SI{4}{\milli\second}$ are highlighted in \cref{fig:tomography}.

We attribute deviation from the exact dynamics to motional dephasing, Trotter error and measurement imperfection, which are analysed in \cref{appendix:error}.
In summary, motional dephasing occurs at a rate of $\gamma_\phi=\SI{18}{\per\second}$, which damps the coherent oscillation.
Trotter error arises from both the higher-order terms in \cref{eq:cosine_evo} and from the phase increment with static laser detuning.
Measurement imperfection is dominated by shot noise, which manifests in the linear fits used to determine $\langle x\rangle$ by \cref{eq:xexpect}.
Shot noise also manifests in the plots of $P(x)$ in \cref{fig:symmetric}c and d.

\subsection{Asymmetric double wells}

We demonstrate the programmability of our scheme by simulating several asymmetric double-well potentials and different initial conditions. 
To do so, we tune the parameters in \cref{eq:cosine_evo}, which necessitates the implementation of $R_x(\varphi)$, which was not needed for the symmetric well.
We quantify the asymmetry of our double-well potentials using the ratio of the wells' depths measured from the top of the barrier. That is, the amount of asymmetry is
\begin{equation}
\Xi=1- \frac{V(x_{\mathrm{max}})-V(x_{\mathrm{min},1})}{V(x_{\mathrm{max}})-V(x_{\mathrm{min},2})},
\end{equation}
where $x_{\mathrm{min},1}$ and $x_{\mathrm{min},2}$ are the local minima of the left and right wells, respectively, and $x_{\mathrm{max}}$ is the local maximum between them.
For symmetric double wells, $\Xi=0$, corresponding to setting $\varphi=0$.

In each simulation, we measure $\langle x \rangle$ using 2PFD with $h=0.4$ and with 200 measurements per point.

\begin{figure}
    \centering 
    \includegraphics[width=\columnwidth]{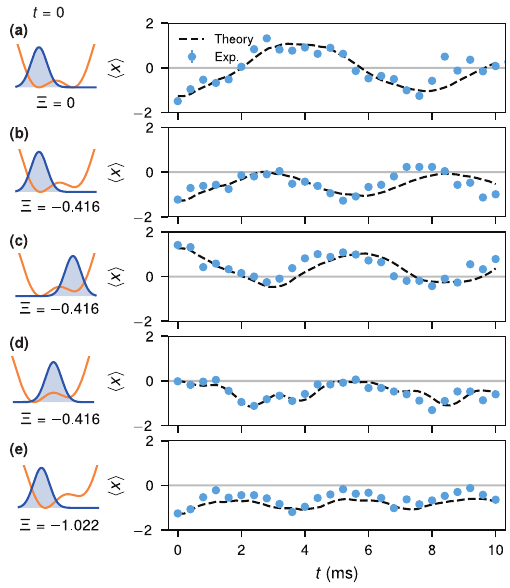}
    \caption{
    \textbf{Programmable anharmonic dynamics.}
    Wavepacket evolution for various potentials and initial states.
    Position expectation values $\langle x \rangle$ are extracted, as functions of time, by 2PFD (\cref{eq:xapprox} with $h=0.4$).
    To engineer asymmetric potentials, we vary the trigonometric-gate angle $\varphi$ from \textbf{(a)}~$\varphi=0$ to \textbf{(b,c,d)}~$\varphi=-\pi/20$ and \textbf{(e)}~$\varphi=-\pi/10$.
    Error bars derived from spin-measurement noise are smaller than the data markers.
    Theory includes post-selection, dephasing, Trotter, and 2PFD errors.
    }
    \label{fig:asymmetric}
\end{figure}

The observed dynamics are shown in \cref{fig:asymmetric}.
We first replicate the symmetric double-well simulation in \cref{fig:symmetric}, but this time estimating the wavepacket position using the 2PFD measurement, showing good agreement with theory.
We then break the symmetry of the double-well potential by varying $\varphi$ in \cref{fig:asymmetric}b--e.
The results demonstrate how the strength of the asymmetry affects the tunnelling dynamics.
In \cref{fig:asymmetric}b,c,d, we introduce an asymmetry of $\Xi=-0.416$.
The asymmetry means that the eigenstates of the potential are no longer symmetric, and so the initial state is not in an equal superposition of the two lowest eigenstates, which has the effect of suppressing the tunnelling.
Whether the wavepacket starts in the left or the right well, the amplitude of the oscillations in $\langle x\rangle$ is smaller than the oscillation for the symmetric double well.
In \cref{fig:asymmetric}d, the wavepacket is initialised in the centre of the asymmetric potential, reflecting the greater overlap of the initial state with the lowest eigenstate.
In \cref{fig:asymmetric}e, we increase the asymmetry to $\Xi=-1.022$, further suppressing the tunnelling.

\subsection{Quartic phase gate from higher-order Fourier synthesis}
\label{sec:gate}

We implement the quartic phase gate $e^{-i\kappa x^4}$ using TGIFS with $N=2$ Fourier terms. 
This unitary follows from \cref{sec:sim_anhm} by setting $V(x)=x^4$ and $\delta=0$.
With $\delta=0$, the phase of the SDD is time-independent, $\zeta=0$, so that the gate depends only on $x$.
To implement a single gate, we set $K=1$ and replace $\Delta t$ with $\kappa$ in \cref{eq:Gc_params_fourier}.

To reduce the number of cosine gates required, we use the Fourier extension (FE)
technique~\cite{Huybrechs2010}.
FE is the same as a standard Fourier series except that the function over the ROI, $[-\tfrac\ell2, \tfrac\ell2]$, is expanded using a frame consisting of periodic functions with period $\Lambda>\ell$.
The FE allows a reduction in the number of terms because it converges faster than the Fourier series when the target function is not periodic over the ROI~\cite{Huybrechs2010}.

For $x^4$, we set $\ell=4$ and $\Lambda=5$.
Because $x^4$ is even, we set $\Phi_n=0$ and determine the FE coefficients $B_n$ by a least-squares fit on the ROI.
With $N=2$, this yields $B_1=-8.59$ and $B_2=3.28$.
We omit the constant term $B_0$, which contributes only a global phase to the gate.

\begin{figure}
    \centering 
    \includegraphics[width=\columnwidth]{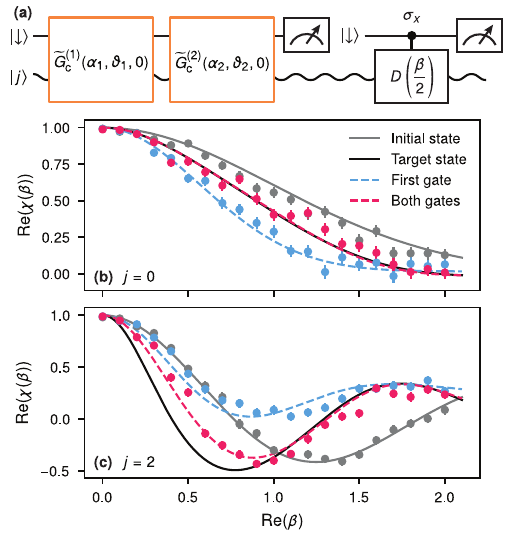}
    \caption{
    \textbf{Quartic phase gate implemented with TGIFS using up to two cosine gates.}
    \textbf{(a)} The circuit used to implement and characterise the quartic phase gate.
    \textbf{(b, c)} Characteristic function measurements of the quartic phase gate applied to the motional ground state $\ket{0}$ (b) and Fock state $\ket{2}$ (c). Only the real part of $\chi$ is measured (the imaginary part is zero by symmetry).
    The state after applying both gates (red markers) is in closer agreement with the target state (black) than applying only a single gate (blue markers).
    Dashed lines show the numerical simulation of the expected characteristic functions, including post-selection and Trotter error.
    }
    \label{fig:gate}
\end{figure}

The circuit used to implement the quartic phase gate is shown in \cref{fig:gate}a.
It consists of the two cosine gates followed by post-selection on the qubit.
We set $\kappa=0.15$ and determine the trigonometric gate arguments from the parameters above and \cref{eq:Gc_params_fourier}. 
We find $\vartheta_1=-1.29$, $\vartheta_2=0.49$, $|\alpha_1| = 0.44$, $|\alpha_2|=2|\alpha_1|$ and  $\varphi_1=\varphi_2=0$.

We characterise the gate performance by measuring $\chi(\beta)$ of the oscillator state after applying the quartic phase gate.
Measurement is performed as described in \cref{methods:meas} and depicted in \cref{fig:gate}a.
The results after applying the gate to Fock states $\ket{0}$ and $\ket{2}$ are shown in \cref{fig:gate}b and c, respectively.
We choose $\ket{2}$ because it tests the gate on a non-classical input.
Initial Fock state preparation is described in \cref{appendix:state_prep}.
For both input states, increasing the number of trigonometric gates in the expansion from zero to one to two brings the output state closer to the target, as expected.

Experimentally measured post-selection acceptance rates for the quartic phase gate acting on $\ket{2}$ are found to be \SI{90}{\percent} for the circuit with only the first gate and \SI{86}{\percent} for the circuit with both gates.
This compares to a theoretical prediction of \SI{94}{\percent} and \SI{93}{\percent}, respectively.
We attribute this discrepancy to stringent conditions placed on the experimental mid-circuit measurement, where we demand that zero photons are detected in order for the state to be counted as dark.

\section{Discussion}

We demonstrate a programmable quantum simulation of anharmonic dynamics using a trapped-ion CV--DV device.
Our approach uses TGIFS to compile a target anharmonic potential into a sequence of non-Gaussian gates, which are decomposed into primitive gates commonly available in current CV--DV devices.
We synthesise a symmetric double-well potential and demonstrate coherent wavepacket dynamics of tunnelling.
To demonstrate programmability, we adjust programmable experimental parameters such as laser phases and pulse durations, synthesising a range of double-well potentials with controllable asymmetries and observing suppression of wavepacket tunnelling.
Using the same circuit structure, we implement the quartic phase gate, $e^{-i\kappa x^4}$, with up to two trigonometric gates, and observe that the measured characteristic function approaches that of the target when the second gate is added.

Our experiment uses the long-lived native coherence of both the DV and CV degrees of freedom in trapped-ion systems.
We use light-atom interactions to implement fully tunable, high-quality CV--DV primitives.
We implement as many as 312 CV--DV entangling gates (state-dependent displacements with magnitude $|\alpha|=\pi/6$ per gate) and observe coherent dynamics up to \SI{15.6}{\milli\second}.

A CV--DV simulator requires far fewer resources than a qubit-only quantum computer to simulate the same double-well dynamics.
To quantify this saving, we compare the resources of our CV--DV simulation with those of a qubit-only simulation with matched error~\cite{Nourse2025}.
We simulate the symmetric double well with the split-operator quantum Fourier transform algorithm on a discrete position grid~\cite{Zalka1998,Kassal2008}, choosing the number of Trotter steps so that the error in $\braket{x}$ matches that of the CV--DV simulation.
On a grid of 32 sites, represented by 5 error-free qubits, this takes 95 Trotter steps comprising a total of $1.2\times10^4$ CNOT gates and $2.1\times10^{5}$ $T$ gates, after synthesising each arbitrary rotation into $T$ gates to precision $10^{-6}$~\cite{Bocharov2015}.
This is compared with one qubit, one bosonic mode and 312 SDDs in the CV--DV simulation.
Thus, CV--DV offers a $2\times10^{3}$-fold improvement in circuit volume, the number of quantum degrees of freedom multiplied by the number of gates.

Looking forward, we envision four complementary routes to expand the range and capability of programmable anharmonic simulations. 

\paragraph*{Increased Fourier terms and Trotter steps.}
Increasing the number of Fourier terms, $N$, would allow TGIFS to simulate increasingly complex anharmonic potentials, while increasing the number of timesteps, $K$, would improve the accuracy of the simulated dynamics.
For example, increasing $N$ would enable the simulation of non-periodic potentials, such as the Morse potential~\cite{Hong2026}, as well as more complex multi-well landscapes.
Overall, the simulation duration scales as $\mathcal{O}(KN^2)$~\cite{TCKS}.
Larger $N$ and $K$ could be achieved through faster gates and longer coherence times, enabled by stronger light-atom coupling and improved motional frequency stabilisation, respectively.
The required circuit depth could also be reduced by efficient compilation techniques, such as Fourier extension.

\paragraph*{Multimode scaling.}
Multimode couplings can be engineered with TGIFS by interleaving SDDs across motional modes.
A Hamiltonian with $N_\mathrm{modes}$ modes, approximated by $N$ Fourier terms, can be simulated using $\mathcal{O}((N_\mathrm{modes} N)^M)$ trigonometric gates per timestep, where $M$ is the maximum number of interacting modes in a single term~\cite{TCKS}.
The multimode scaling becomes polynomial in $N_\mathrm{modes}$ when the interaction is local~\cite{Lloyd1996}, that is, when $M$ is a constant independent of $N_\mathrm{modes}$.
This structure is present in many significant simulation targets, such as anharmonic molecular vibrations where $M$ is typically two to four~\cite{Nielsen1951, Barone2005}.
Trapped ions present a natural route to realising such simulations due to the native coupling of all ions to all modes.
Large Coulomb crystals provide multiple motional modes, with cooling to the motional ground state and crystal stability already demonstrated~\cite{Kiesenhofer2023, Guo2024}; the remaining challenges amount to addressing individual modes with sufficient Rabi rates relative to motional dephasing.

\paragraph*{Time-dependent Hamiltonians.} TGIFS natively supports simulation of time-dependent Hamiltonians.
Time dependence can be implemented by updating the control parameters from one timestep to the next, which would enable quantum simulations of laser-driven chemical reactions~\cite{Shapiro2011quantum}, and chaotic systems~\cite{Gardniner1997}.

\paragraph*{Engineered dissipation.} 
TGIFS is, in principle, compatible with the existing framework for simulating open quantum systems by controlled noise injection~\cite{Sun2025, So2024, Olaya-Agudelo2025, Navickas2025}. This opens opportunities to simulate dissipative anharmonic dynamics in molecular systems such as proton transfer in the condensed phase~\cite{Craig2007proton,vanHaeften2023propagating}, energy transport~\cite{Guo2015tuning,Chen2020steadystate} and bond breaking~\cite{Gao1998lindblad,Gao2001Disspiative}.

Overall, our experiment validates TGIFS as a compiler for simulating anharmonic potentials, provides an experimental demonstration of BQSP, and establishes BQSP as a practical framework for implementing programmable non-Gaussian operations.
Extending our approach as detailed above would broaden the class of anharmonic dynamics accessible to CV--DV simulators, enabling high-throughput simulations of physical models spanning applications from quantum chemistry to condensed-matter and high-energy physics.

\begin{acknowledgments}

We thank Jungsoo Hong and Julian Jee for valuable discussions which have helped to improve this work.
We were supported by the U.S. Office of Naval Research Global (N62909-24-1-2083), the U.S. Air Force Office of Scientific Research (FA2386-23-1-4062), the Wellcome Leap Quantum for Bio program, the Australian Research Council (FT220100359, FT230100653), Lockheed Martin, the Sydney Quantum Academy (MJM, PN), the University of Sydney Postgraduate Award scholarship (VGM), the Australian Government Research Training Program (FS), the Sydney Horizon Fellowship (TRT), and H.\ and A.\ Harley.

\end{acknowledgments}

\section*{Author Contributions Statement}

CM, TC, KS, VGM, CHV, IK, TRT conceived the idea. 
CM, FS, TC, KS developed the experimental protocols. 
CM performed the experiment. CM, TC analysed the data. 
CM, FS, MJM, VGM, CHV, PN, HKC, TRT contributed to the experimental apparatus, and TC, KS, HLN provided theoretical support. 
IK and TRT supervised the project. 
CM, TC, KS, IK, TRT wrote the manuscript. All authors provided suggestions for the experiment, discussed the results, and contributed to the manuscript.

\section*{Competing Interests Statement}

The authors declare no competing interests.

\section*{Data Availability}

A repository containing the experimental data presented in this article is available at \inlineref{dataset}.

\section*{Appendices}

\appendix

\section{Generating trigonometric gates}
\label{appendix:trig_gates}

Here, we provide the derivation of the operator $Q(\alpha, \vartheta, \varphi)$ constructed with a BQSP sequence consisting of a single-qubit rotation and two state-dependent displacement operators. We begin with the simple case of $\varphi=0$, which is used to synthesise the symmetric double-well potential. 
We begin the derivation by rewriting \cref{eq:Q_def} as
\begin{align}
    Q(\alpha, \vartheta, 0) &= D(-\sigma_x\alpha)e^{i\vartheta\sigma_z/2}D(\sigma_x\alpha)\\
    &= \exp\left(\tfrac12 i\vartheta D(-\sigma_x\alpha)\sigma_zD(\sigma_x\alpha)\right).
    \label{eq:D_in_exp}
\end{align}
where we use the fact that $D(\sigma_x\alpha)$ is unitary and $D(-\sigma_x\alpha)=D^\dagger(\sigma_x\alpha)$.

\Cref{eq:D_in_exp} can be simplified using the identity
\begin{equation}
e^{i\Theta \sigma_x}\sigma_ze^{-i\Theta \sigma_x}
= \cos(2\Theta)\sigma_z + \sin(2\Theta)\sigma_y,
\label{eq:unitaryID}
\end{equation}
which holds when $\Theta$ is Hermitian and $[\Theta, \sigma_z]=0$.
In our case, $\Theta = i\left(\alpha a^\dagger - \alpha^*a\right)=\sqrt{2}|\alpha|x_\zeta$, which yields
\begin{multline}
    Q(\alpha, \vartheta,0) = \exp\big(\tfrac12 i\vartheta\big(\sigma_z\cos(2\sqrt{2}|\alpha| x_\zeta) \\
    \hspace{2cm}+\sigma_y\sin(2\sqrt{2}|\alpha| x_\zeta)\big)\big),
    \label{eq:Q_varphi0}
\end{multline}
which is then used to synthesise a trigonometric gate $\widetilde{G}_\mathrm{c}(\alpha, \vartheta, 0)$ to simulate anharmonic dynamics in a symmetric double-well potential. 

This approach is straightforwardly extended to asymmetric double wells. 
Here, we have a nonzero $\varphi$ in the BQSP operation $Q(\alpha, \vartheta, \varphi)$ defined in \cref{eq:Q_def}, 
\begin{align}
    Q(\alpha, \vartheta, \varphi) =&  D(-\sigma_x\alpha) R_\varphi(-\vartheta) D(\sigma_x\alpha),\nonumber\\
    =& D(-\sigma_x\alpha) R_x(-\varphi)R_z(-\vartheta)R_x(\varphi) D(\sigma_x\alpha).\label{eq:Q_varphi1}
\end{align}
Because the $R_x$ rotations commute with $D(\sigma_x\alpha)$, their orders can be swapped to rewrite \cref{eq:Q_varphi1} as 
\begin{equation}
Q(\alpha, \vartheta, \varphi) = R_x(-\varphi)Q(\alpha, \vartheta, 0)R_x(\varphi).
\end{equation}
As the rotation operators are unitary, they can be moved inside the exponent of $Q(\alpha, \vartheta, 0)$ (\cref{eq:Q_varphi0}), 
\begin{multline}
Q(\alpha, \vartheta, \varphi) = \exp\big(\tfrac12 i\vartheta\big(R_x(-\varphi)\sigma_zR_x(\varphi)\cos(2\sqrt{2}|\alpha| x_\zeta) \\
+ R_x(-\varphi)\sigma_yR_x(\varphi)\sin(2\sqrt{2}|\alpha|x_\zeta)\big)\big).
\label{eq:Q_varphi2}
\end{multline}
Here, we apply \cref{eq:unitaryID} to $R_x(-\varphi)\sigma_zR_x(\varphi)$ inside the exponent, with $\Theta=\varphi/2$.
We use a similar identity for $R_x(-\varphi)\sigma_yR_x(\varphi)$.
Finally, gathering trigonometric terms gives \cref{eq:Q}.

\section{State preparation}
\label{appendix:state_prep}

As part of the experimental initialisation, we prepare the ion's spin and motional ground states in three steps.

First, we apply Doppler cooling to the ion using a \SI{369.5}{\nano\meter} laser that is red-detuned from the Yb$^+$ ion's  \ce{^2S_{1/2}} $\rightarrow$ \ce{^2P_{1/2}} transition.
This laser is aligned with components along all trap axes, cooling all motional modes.

Second, radial modes are cooled to their motional ground states with pulsed sideband cooling~\cite{Leibfried2003}, using the same \SI{355}{\nano\meter} laser described in \cref{sec:methods}.
We cool both radial modes; the axial mode is not cooled as it is inaccessible due to our laser beam geometry. 
We typically reach an average phonon number of $\bar{n} = 0.04$, verified by sideband thermometry~\cite{Monroe1995}.

Third, the spin is prepared in the ground state by optically pumping on resonance with the \ce{^2S_{1/2}}~$\ket{F=1}\rightarrow$~\ce{^2P_{1/2}}~$\ket{F=1}$ transition. 

When the motional state $\ket{2}$ is required, as in \cref{sec:gate}, we prepare it as follows.
First, we prepare $\ket{1}$ by a BSB $\pi$ pulse followed by a carrier $\pi$ pulse.
The sequence is heralded by a mid-circuit measurement: when the measurement is bright, the result is discarded.
We repeat this sequence with a reduced BSB $\pi$-time to map from $\ket{1}$ to $\ket{2}$.

\section{Qubit measurement}
\label{appendix:measurement}

We determine the state of the qubit by fluorescence measurement of the $\ket{\uparrow}$ state by driving the \ce{^2S_{1/2}}~$\ket{F=1}\rightarrow$~\ce{^2P_{1/2}}~$\ket{F=0}$ transition.
Scattered light is collected by a high-numerical-aperture imaging system and focused onto an avalanche photodiode. The qubit state is inferred from the number of scattered photons via thresholding: photon counts higher than a pre-calibrated threshold indicate the  $\ket{\uparrow}$ (bright) state, while counts lower than the threshold give the $\ket{\downarrow}$ (dark) state.
We use a measurement duration of \SI{250}{\micro\second} and a photon threshold of two.

We post-select results based on mid-circuit measurements that yield binary outcomes; if a bright state is detected at any mid-circuit measurement, the result is discarded.
Only keeping the dark state suppresses photon scattering by the ion during the mid-circuit measurements, thereby avoiding the effects of photon recoil on the motional state.
Mid-circuit measurements are performed for \SI{50}{\micro\second}, with a threshold of one photon.
The outcome is stored in a buffer memory of our experimental control hardware and read out at the end of the entire sequence (after the final measurement). 

\begin{figure}
    \centering
    \includegraphics[width=\columnwidth]{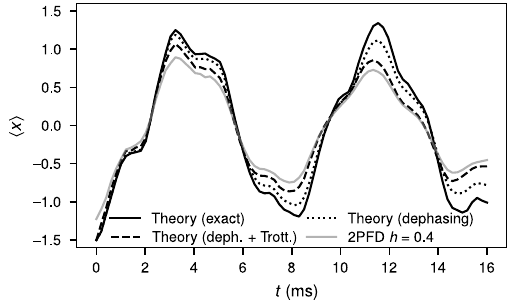}
    \caption{
    \textbf{Errors due to dephasing, Trotter error and 2PFD measurement.}
    Traces show the effect as additional error sources are added to the theoretical model of our simulation of a symmetric double well.
    Shown are: exact theoretical dynamics described by $H_\text{sim}$, theoretical dynamics including dephasing, theoretical dynamics including both dephasing and Trotter error, and theoretical dynamics with dephasing, Trotter error and 2PFD readout error for $h=0.4$.
    }
    \label{fig:errors}
\end{figure}

\begin{table}[b]
  \caption{
  \label{tab:errors}
   \textbf{SRMSE due to dephasing, Trotter error and 2PFD measurement}, corresponding to the numerical results in \cref{fig:errors}.
   The first numeric column reports the incremental SRMSE obtained when successive error mechanisms are included (e.g., theory with dephasing versus the exact solution; theory with dephasing and Trotterisation versus theory with dephasing; etc.).
   The second numeric column reports the SRMSE of each model when compared to the exact solution.
  }
  \begin{ruledtabular} 
    \begin{tabular}{lll}
      Source              & SRMSE (\si{\percent})   & Cumulative SRMSE (\si{\percent}) \\
      \colrule
      Dephasing           & 13.6  & 13.6 \\
      Trotter & 19.2  & 29.8 \\
      2PFD $(h=0.4)$   & 3.7  & 31.9
    \end{tabular}
  \end{ruledtabular}
\end{table}

\section{Sources of error}
\label{appendix:error}

Three main error sources cause deviation of experimental results from the exact dynamics: experimental noise (which is dominated by motional dephasing), Trotter error, and errors associated with the two-point finite difference (2PFD) measurement scheme.

Dephasing occurs due to amplitude fluctuations of the radio-frequency signal that generates the trapping potential~\cite{Leibfried2003, Valahu2023}.
To combat this, the amplitude is locked to a reference voltage via a proportional-integral-derivative feedback loop~\cite{MacDonell2023} and the remaining slow fluctuations are partially mitigated by periodically recalibrating the motional frequencies during each experimental run.
Dephasing is modelled by solving the Lindblad master equation for $H_\mathrm{sim}$ with the Lindblad operator $L_{\gamma_\phi}=\sqrt{\gamma_\phi}a^\dagger a$, where $\gamma_\phi=\SI{18}{\per\second}$ is the measured dephasing rate~\cite{Matsos2024}.

Trotter error associated with our approximation of $\widetilde{G}_\mathrm{c}$ in \cref{eq:cosine_evo} leads to further deviation from the target dynamics.
The error can be separated into two components.
The first is the $\mathcal{O}(\vartheta^2)$ term in \cref{eq:cosine_evo}, which could be reduced by a higher-order Trotter expansion or a shorter timestep~\cite{TCKS}.
Spin flips associated with this Trotter error are mitigated by the mid-circuit measurement, as described in \cref{methods:time_evo}.
The second error arises from setting a static laser detuning $\delta_\mathrm{L}$ in $H_\text{SDD}$ (\cref{eq:H_sdd_main}) to implement the phase increment in $\widetilde{G}_\mathrm{c}$.
This error becomes negligible in the limit of $|\delta_\mathrm{L}|\tau\ll1$.
Trotter error is included in the simulation by solving the Lindblad master equation with dephasing Lindbladian operator $L_{\gamma_\phi}$ for the complete pulse sequence of SDD and SQR primitives at each timestep.

The 2PFD slope measurement scheme used to extract the data in \cref{fig:asymmetric} introduces an error that depends on $h$, the distance of the sample point from the origin.
The scheme demands that $h$ is chosen to be sufficiently small, so that the second measurement point is taken in the regime where $\Im[\chi(\beta)]$ remains linear.

\Cref{fig:errors} plots dynamics with these errors and \Cref{tab:errors} quantifies the scaled root-mean-square error (SRMSE) for each error source.
For two series of length $N_\mathrm{t}$, $a=\{a_k\}$ and $b=\{b_k\}$, the SRMSE is
\begin{equation}
\epsilon(a, b) = \sqrt{\frac{\sum_{k=1}^{N_\mathrm{t}} (a_k - b_k)^2}{\sum_{k=1}^{N_\mathrm{t}} a_k^2}}.
\end{equation}
The experimental data deviate from the exact dynamics largely because of the dephasing and Trotter errors, with 2PFD errors being relatively minor at our choice of $h=0.4$.

\begin{figure}
    \centering
    \vspace{4mm}
    \includegraphics[width=\columnwidth]{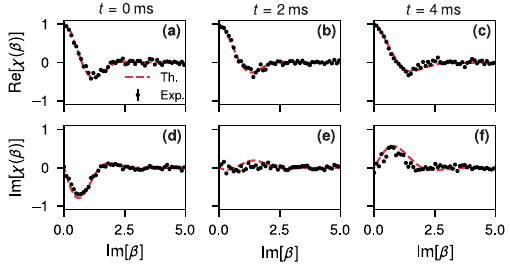}
    \caption{
    \textbf{Measured characteristic function} used to construct the probability densities in \cref{fig:symmetric}.
    \textbf{(a--c)} The real parts of $\chi(\beta)$ at $t=\SI{0}{ms}$, \SI{2}{\milli\second} and \SI{4}{\milli\second}, respectively.
    \textbf{\mbox{(d--f)}} Corresponding imaginary parts.
    Theoretical predictions (including dephasing and Trotter error) are given for comparison.
    The error bars correspond to measurement projection noise, and are smaller than the points.
    We measure $\chi(\beta)$ at 50 points in the range $\Im[\beta]\in[0,5]$ in randomised order, with 500 measurements per point.
    }
    \label{fig:probs_cf}
\end{figure}

\begin{figure*}
    \centering
   \includegraphics[width=\textwidth]{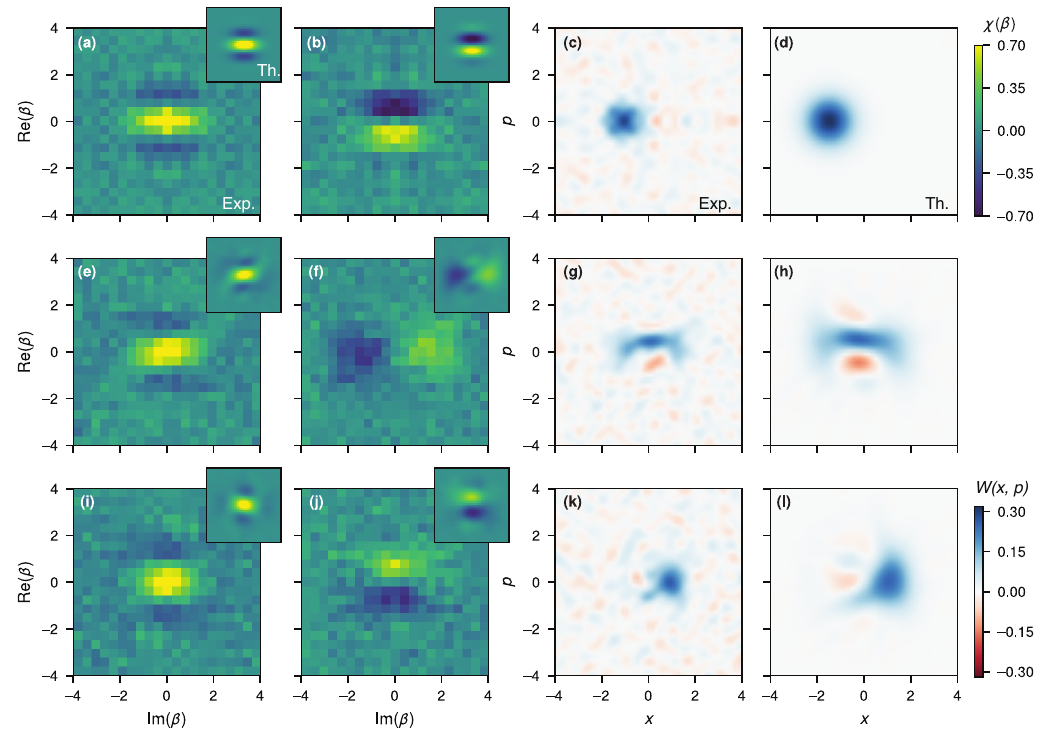}
    \caption{
    \textbf{Two-dimensional scans of the characteristic and reconstructed Wigner functions for the symmetric double well,} supplementing the data in \cref{fig:symmetric}.
    \textbf{(a)} $\Re[\chi(\beta)]$ at $t=\SI{0}{\milli\second}$, measured on a $21\times21$ grid.
    Inset: theoretical $\Re[\chi(\beta)]$ with dephasing and Trotter error.
    \textbf{(b)} The corresponding imaginary part of $\chi(\beta)$.
    \textbf{(c)} Measured Wigner function $W(x,p)$ at $t=\SI{0}{\milli\second}$.
    \textbf{(d)} Theoretical $W(x,p)$ with dephasing and Trotter error.
    The next two rows show the same plots at $t=\SI{2}{\milli\second}$ (\textbf{(e)--(h)}) and $t=\SI{4}{\milli\second}$ (\textbf{(i)--(l)}).
    }
    \label{fig:wigners}
\end{figure*}

\section{Motional state measurement}
\label{appendix:motional_meas}

In this section, we present further measurements of the oscillator state in the symmetric double-well case.

Figure~\ref{fig:probs_cf} shows the characteristic function $\chi(\beta)$ measurements that are used to construct the probability distributions in \cref{fig:symmetric}.

In addition, we perform a full two-dimensional scan of $\chi(\beta)$ at $t=\SI{0}{\milli\second}$, \SI{2}{\milli\second}, and \SI{4}{\milli\second} (see \cref{fig:wigners}).
We measure $\chi(\beta)$ in the range $\Re[\beta]\in[-4,4]$ (21 points) and $\Im[\beta]\in[0,4]$ (11 points), and use the symmetry of $\chi$ to determine the unmeasured values.
At $t=0$, measuring only the real quadrant is sufficient to determine $\chi$ because the prepared state is even in momentum, i.e., $\chi(\beta)= \chi(\beta^{*})$. 
The data are collected in a random order with 250 measurements per point.
Wigner functions are obtained from the Fourier transform of $\chi(\beta)$ using \cref{eq:WignerFT}.
When calculating $W(x,p)$, $\chi(\beta)$ data are padded with zeros (out to $\left|\Re[\beta]\right|<10$ and $\left|\Im[\beta]\right|<10$), and the background is subtracted from each $\chi(\beta)$ (mean value $(-5.0+0.1i)\times10^{-3}$)~\cite{Valahu2023}.

\section{Implementing the Kerr gate}
\label{appendix:Kerr}

The Kerr gate can be constructed from a series of quartic phase gates as follows.
We write $x_\zeta^4$ in terms of creation and annihilation operators
\begin{align}
x_\zeta^4 = \frac{1}{4}\Big( & e^{4i\zeta} a^{\dagger 4} + e^{-4i\zeta} a^{4} \nonumber \\
& + 2e^{2i\zeta}\left(2a^{\dagger 3}a + 3a^{\dagger 2}\right) + 2e^{-2i\zeta}\left(2a^{\dagger}a^{3} + 3a^{2}\right) \nonumber \\
& + 6a^{\dagger 2}a^{2} + 12a^{\dagger}a + 3 \,\Big),
\end{align}
and note that the terms with $\zeta$ dependence can be cancelled when summed over multiple quadrature angles, for example
\begin{equation}
x_0^4 + x_{\pi/3}^4 + x_{2\pi/3}^4 = \frac34( 6{a^\dagger}^2 a^2 + 12a^\dagger a +3 ),
\end{equation}
where the three terms on the right-hand side correspond to the Kerr nonlinearity, a frequency shift and a constant, respectively.
We can therefore write, using the Lie--Trotter product,
\begin{equation}
\prod_{j=0}^2e^{-i\tfrac{2\kappa}9 x_{j\pi/3}^4} \approx \exp\left(-i\kappa (a^{\dagger2}a^2 + 2a^\dagger a + \tfrac{1}{2})\right),
\end{equation}
where the approximation holds to first order in $\kappa$.
The rotation in phase space caused by the $a^\dagger a$ term in the exponent can be accounted for by adjusting the phase in subsequent pulses.
The global phase introduced by the constant term can be ignored.
The remaining ${a^\dagger}^2a^2$ interaction yields the desired Kerr gate.

\bibliographystyle{apsrev4-2}
\bibliography{bib}

\end{document}